\documentclass{emulateapj}
\usepackage{graphicx}
\begin{document}

\def\etal{et al.\ \rm}
\def\ba{\begin{eqnarray}}
\def\ea{\end{eqnarray}}
\def\etal{et al.\ \rm}

\title{Runaway accretion of metals from compact debris disks onto 
white dwarfs.}

\author{Roman R. Rafikov\altaffilmark{1,2}}
\altaffiltext{1}{Department of Astrophysical Sciences, 
Princeton University, Ivy Lane, Princeton, NJ 08540; 
rrr@astro.princeton.edu}
\altaffiltext{2}{Sloan Fellow}


\begin{abstract}
It was recently proposed that metal-rich white dwarfs (WDs) 
accrete their metals from compact debris disks found to exist 
around more than a dozen of them. At the same time, elemental
abundances measured in atmospheres of some WDs imply vigorous 
metal accretion at rates up to $10^{11}$ g s$^{-1}$, far 
in excess of what can be supplied solely by Poynting-Robertson 
drag acting on such debris disks. To explain this observation 
we propose a model, in which rapid transport 
of metals from the disk onto the WD naturally results from 
interaction between this particulate disk and spatially 
coexisting disk of metallic gas. The latter is fed 
by evaporation of debris particles at the sublimation radius
located at several tens of WD radii. Because of pressure 
support gaseous disk orbits WD slower than particulate 
disk. Resultant azimuthal drift between them at speed 
$\lesssim 1$ m s$^{-1}$ causes aerodynamic drag on the disk of solids 
and drives inward migration of its constituent particles. 
Upon reaching the sublimation radius particles evaporate, 
enhancing the density of metallic gaseous disk and 
leading to positive feedback. Under favorable circumstances 
(low viscosity in the disk of metallic gas and efficient 
aerodynamic coupling between the disks) system evolves in a 
runaway fashion, 
destroying debris disk on time scale of $\sim 10^5$ yr, 
and giving rise to high metal accretion rates 
up to $\dot M_Z\sim 10^{10}-10^{11}$ g s$^{-1}$,
in agreement with observations. 
\end{abstract}



\section{Introduction.}  
\label{sect:intro}


Recent detections of near-infrared excesses around a number 
of metal-rich white dwarfs (WD) imply existence of warm 
circumstellar material reprocessing stellar radiation 
(Zuckerman \& Becklin 1987; Graham \etal 1990; Farihi \etal 2010).
Spectral modeling suggests that this material 
resides in an extended, compact, 
optically thick and geometrically thin disk (Jura 2003; 
Jura \etal 2007), similar to the Saturn's rings (Cuzzi \etal 2010). 
Disks have inner edges at several tens of the WD radii 
$R_\star$, roughly consistent with being set by particle 
sublimation at these locations.
Their outer radii lie close to the Roche radius of the WD
$R_R\sim 1$ R$_\odot$, supporting the suggestion by Jura 
(2003) that such compact debris disks are produced by tidal 
disruption of asteroid-like bodies scattered into low-periastron 
orbits by gravitational perturbations of massive 
planets, which have survived the AGB phase of stellar 
evolution. 

Availability of large reservoir of high-Z elements in the 
form of debris disk in the immediate vicinity of some WDs
naturally led to the suggestion (Jura 2003) that metal 
enrichment of these stars is caused by accretion from 
such disks. This scenario provides a promising alternative 
to the previously proposed {\it interstellar} accretion model of 
metal pollution of WDs (Dupuis \etal 1993), which is not 
consistent with observations of WDs with He atmospheres.

Theoretical estimates of settling time of heavy elements 
in WDs imply that their observed  atmospheric abundances 
can be maintained against gravitational settling
by accretion of metals at rates $\dot M_Z\sim 10^6-10^{11}$ 
g s$^{-1}$ (Farihi \etal 2009, 2010). If the {\it circumstellar} 
accretion hypothesis is correct, an evolving disk of debris 
must be able to supply such $\dot M_Z$ to the WD.

The actual transfer of metals from the disk of solids 
truncated at sublimation radius $R_s$ to the WD 
must be accomplished in this picture via the gas disk 
extending from the WD surface to $R_s$ and beyond. 
Observational evidence of such gaseous component around 
several metal-rich WDs hosting compact debris disks has been
found by G\"ansicke \etal (2006, 2007, 2008) in the form of
double-peaked emission lines of Ca II and Fe II. These
spectroscopic signatures are naturally explained as
arising in a disk of metallic gas (no H or He emission lines 
have been detected around these WDs) in Keplerian rotation 
around WDs and spatially coincident with dusty disks 
(Melis \etal 2010).

Even though metals are passed to the WD through the gaseous disk 
the rate of mass transfer $\dot M_Z$ is ultimately controlled 
by evolution of the particulate disk. By analogy with 
Saturn's rings one expects that the collisional viscosity 
in debris disk is too low to drive the non-negligible $\dot M_Z$.
However, radiation of the WD can be quite important and Rafikov 
(2011; hereafter R11) demonstrated that Poynting-Robertson 
(PR) drag on the disk naturally drives mass accretion at the rate 
$\dot M_{PR}\gtrsim 10^8$ g s$^{-1}$. While this $\dot M_Z$ is 
high it still falls short of explaining 
the highest observed  $\dot M_Z\sim 10^{10}-10^{11}$ 
g s$^{-1}$. 

Here we propose a new picture of the debris  
disk evolution, which naturally combines several 
physical ingredients present in close vicinity of 
disk-bearing WDs.


\section{Description of the model.}
\label{sect:model}

\begin{figure}
\plotone{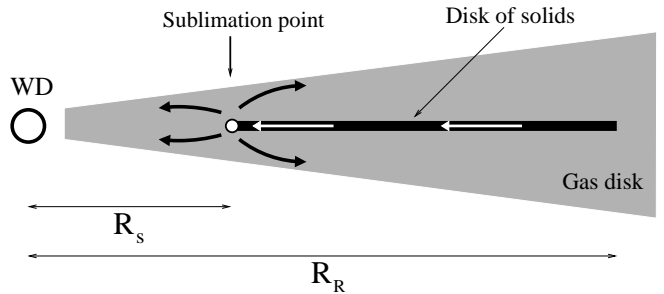}
\caption{
Schematic picture of two interacting disks (black - 
particulate, gray - gaseous) around the WD. Disk of solid
particles is confined between the sublimation radius $R_s$ 
(where solids evaporate feeding gas disk) and Roche radius 
$R_R$. Flow of high-Z material is indicated by arrows 
(white - solid particles, black - gas). Inward migration of 
solids is induced by drag exerted on them by the 
gas disk, which orbits WD slower than the disk of particles.
\label{fig:f1}}
\end{figure}

Our model of the WD-disk system is described below 
and is schematically illustrated in Figure \ref{fig:f1}.
We assume a disk of solid particles to extend from 
the Roche radius $R_R$ all the way in to the sublimation 
radius
\ba
R_s\equiv \frac{\sqrt{\epsilon} R_\star}{2}
\left(\frac{T_\star}{T_s}\right)^{2}\approx 22~
\sqrt{\epsilon} 
R_\star T_{\star,4}^2
\left(\frac{1500\mbox{K}}{T_s}\right)^2,
\label{eq:R_S}
\ea
where $T_s$ is the sublimation temperature of solid 
particles (which we take to be $\sim 1500$ K if particles 
are silicate), $T_{\star,4}\equiv T_\star/(10^4$ K) is 
the normalized stellar temperature $T_\star$, $R_\star$ is 
the WD radius, and $\epsilon$ is the ratio
of particle emissivities for starlight and for its own 
thermal radiation (in the following we assume macroscopic 
particles and set $\epsilon=1$). For $R_\star\approx 0.01R_\odot$ 
(Ehrenreich \etal 2011) one finds $R_s\approx 0.2$ R$_\odot$, 
in agreement with observations (Jura \etal 2007, 2009).

At present the size of particles $a$ constituting the disk 
is rather poorly constrained. Graham \etal (1990) 
suggested $a\approx 10-100$ cm. At the same time high-resolution 
infrared spectroscopy with IRS onboard of 
{\it Spitzer} reveals strong $10~\mu$m emission feature
in disk spectra indicative of the existence of a population of 
small, micron-sized silicate dust particles
(Jura \etal 2009). But the fraction of total disk mass locked 
up in such fine dust is unknown. The actual particle size is 
not very important for our treatment as long as the disk is optically 
thick. Then it is similar to Saturn's rings and most 
likely behaves as a granular flow. In the following we will 
treat particulate disk as if it were a solid plate. 

Solid particles brought to $R_s$ sublimate feeding a disk of 
metallic gas at this location. Viscous torques cause it
to spread all the way to the WD surface providing means of 
metal transport from $R_s$ to the star. However, because of 
angular momentum conservation part of the newly produced 
metallic gas has to move {\it outwards} of $R_s$ 
(Lynden-Bell \& Pringle 1974) naturally explaining 
the existence of gaseous disks spatially coexisting with 
dusty debris disks in some systems (Melis \etal 2010). 
Jura (2008) proposed collisional sputtering of small 
asteroids as another source of metallic gas; for the sake of 
clarity we do not consider this mechanism here.

We emphasize here that the two disks have an overlap in 
{\it radial} distance, while in vertical 
direction gaseous disk is much more extended than the disk 
of particles, see Figure \ref{fig:f1}. The latter must be 
very thin because 
inelastic collisions between particles rapidly damp any 
vertical random motions.

External gas disk outside $R_s$ has temperature 
different from (higher than) that of the dust disk at the 
same radii (Melis \etal 2010) because of the different 
balance of heating and cooling for the two disks. This keeps 
gas temperature $T_g$ above the sublimation point even 
outside $R_s$, although some condensation of metallic 
gas on the particle surfaces may be happening there 
(we neglect it in this work).

Simultaneous existence of the two disks of high-Z elements 
in different phases drives their mutual evolution in the 
following way. 
Any coupling between the outer portion of the gas disk and
particulate disk acts to transfer 
angular momentum from the faster rotating particulate disk
to the slower spinning gaseous disk, causing inward motion 
of particles in the disk of solids. If the coupling is strong 
enough positive feedback becomes possible in the system: 
increasing mass of gaseous disk leads to the increase of 
$\dot M_Z$ through the disk of solids (as described below), 
which in turn reinforces evaporation at $R_s$ and increases 
gaseous mass even further.

Coupling between the gaseous and particulate high-Z disks 
arises because gaseous disk orbits WD at angular speed 
$\Omega_g$ slightly lower than the Keplerian speed $\Omega_K$
at which the disk of solid particles rotates. This difference 
is caused by the pressure support present in gaseous disk, 
$\Omega_g-\Omega_K\approx (2\Omega_K r\rho)^{-1}\partial P/\partial r$,
and is known to cause a variety of important effects in 
protoplanetary disks, such as inward migration of solids 
(Weidenschilling 1977). Relative azimuthal velocity between 
the gaseous and particulate disks at distance $r$ from the 
WD is 
\ba
v_{\varphi}=\eta c_s\frac{c_s}{\Omega_K r}\approx 10^2\mbox{cm s}^{-1}
\frac{T_{g,3}}{\mu_{28}}\left(M_{\star,1}^{-1}\frac{r}{0.2R_\odot}
\right)^{1/2},
\label{eq:v_rel}
\ea
where $\eta\sim 1$ is a constant, 
$M_{\star,1}\equiv M_\star/M_\odot$ is the normalized 
WD mass $M_\star$, 
$\mu_{28}$ is the mean molecular weight of the metallic 
gas normalized by $28m_p$ (value of $\mu$ for pure Si), 
$T_{g,3}\equiv T_g/(10^3$ K) is the normalized gas temperature 
$T_g$, and $c_s\approx 0.5$ km s$^{-1}
(T_{g,3}/\mu_{28})^{1/2}$ is the gas sound speed (clearly 
$v_{\varphi}\ll c_s$).

This azimuthal drift gives rise to aerodynamic drag between 
the disks. Azimuthal drag force $f_\varphi$ per unit surface 
area of a particulate disk with surface mass density $\Sigma_d$ 
causes inward radial migration of the disk material at speed 
$v_{r}=2f_\varphi/(\Omega_K\Sigma_d)$. This inward  
particle drift gives rise to mass transport at the rate 
\ba
\dot M_Z=2\pi r v_{r}\Sigma_d=\frac{4\pi r f_\varphi}{\Omega_K}.
\label{eq:dotMgen}
\ea
Note that if $f_\varphi$ is independent of $\Sigma_d$ then 
the same is true for $\dot M_Z$. 

External force $f_\varphi$ per unit area can be generally 
represented in the form
\ba
f_\varphi=A\Sigma_g+B,
\label{eq:f_e_decomp}
\ea
where the first term describes the coupling between the 
gas disk with surface density $\Sigma_g$ and particulate disk, 
and constant $A$ determines the strength of coupling. It is 
natural to expect that drag scales with gas density $\Sigma_g$;
this expectation is confirmed in \S\ref{sect:aero}.

Second term $B$ represents forces 
acting even in the absence of coupling to gas disk. 
PR drag is an example of such force and one can easily show 
(R11) that 
for PR drag $B_{PR}\approx\alpha\Omega_K L_\star \phi_r/(4\pi r c^2)$, 
where $c$ is the speed of light, $\phi_r\sim 1$ is the efficiency 
of radiative momentum absorption by the disk surface, and 
$\alpha\approx (4/3\pi)(R_\star/r)$ 
(Friedjung 1985) is the incidence angle of stellar radiation. 
From equation (\ref{eq:dotMgen}) the rate of mass transport 
due to PR drag alone (when $A=0$) at $R_s$ is (R11)
\ba
\dot M_{PR}&=&\frac{4\pi R_s B_{PR}}{\Omega_K}=\frac{4\phi_r}{3\pi}
\frac{R_\star}{R_s}\frac{L_\star}{c^2}
\label{eq:Mdot_PR}\\
&\approx & 10^8\mbox{g s}^{-1}\phi_r\frac{L_\star}
{10^{-3}L_\odot}\frac{20}{R_s/R_\star}.
\nonumber
\ea  
In the following we will assume that $B=B_{PR}$.


\section{Coupled evolution of the particulate and gaseous disks.}
\label{sect:coupled}


Particle sublimation at $R_s$ increases the mass of gaseous disk
at the rate $\dot M_Z$. Assuming that surface density of gaseous disk 
$\Sigma_g$ varies on scale $\sim R_s$ we can write that sublimation 
increases $\Sigma_g$ at the rate $\dot\Sigma_+\sim \dot M/(\pi R_s^2)
=4f_\varphi/(\Omega_K R_s)$.
At the same time, viscous spreading reduces $\Sigma_g$ at the rate
$\dot\Sigma_-\sim \Sigma_g/t_\nu$, where $t_\nu$ is the 
characteristic viscous time in the disk 
\ba
t_\nu\sim \frac{R_s^2}{\nu}\approx 10~\mbox{yr}~\alpha^{-1}
\frac{\mu_{28}}{T_{g,3}}
\left(M_{\star,1}
\frac{R_s}{0.2~\mbox{R}_\odot}\right)^{1/2},
\label{eq:t_nu}
\ea
assuming $\alpha$-parametrization of viscosity 
$\nu=\alpha c_s^2/\Omega_K$ (Shakura \& Sunyaev 1973).
This timescale can be very short if $\alpha$ is not 
very small. 

We can now describe the evolution of $\Sigma_g$ in the 
vicinity of $R_s$ with the following heuristic equation:
\ba
\frac{\partial \Sigma_g}{\partial t}=\dot\Sigma_+-\dot\Sigma_-=
\frac{4}{\Omega_K R_s}(A\Sigma_g+B)-\frac{\Sigma_g}{t_\nu}.
\label{eq:heuristic}
\ea
A solution of this equation satisfying the initial condition
$\Sigma_g(t=0)=0$ (no gas disk initially) is
\ba
\Sigma_g(t)&=&\frac{\dot M_{PR}t_s}{\pi R_s^2}
\left(1-\frac{t_s}{t_\nu}\right)^{-1}\nonumber\\
&\times &\left\{\exp\left[\frac{t}{t_s}
\left(1-\frac{t_s}{t_\nu}\right)\right]-1\right\},
\label{eq:sol}
\ea
where we used equation (\ref{eq:Mdot_PR}) and 
\ba
t_s=\frac{\Omega_K R_s}{4A},
\label{eq:tau_s}
\ea
is the {\it sublimation time} $\Sigma_g/\dot\Sigma_+$
on which the mass of gaseous disk increases if $B_{PR}=0$.
From equations (\ref{eq:dotMgen}), (\ref{eq:f_e_decomp}) 
and (\ref{eq:sol}) the rate at which particles sublimate is
\ba
\frac{\dot M_Z}{\dot M_{PR}}=
\left(1-\frac{t_s}{t_\nu}\right)^{-1}
\left\{\exp\left[\frac{t}{t_s}
\left(1-\frac{t_s}{t_\nu}\right)\right]-\frac{t_s}{t_\nu}\right\},
\label{eq:MZsol}
\ea
and the total mass lost by the debris disk to sublimation $M_s$ 
can be easily obtained by integrating this expression.

\begin{figure}
\plotone{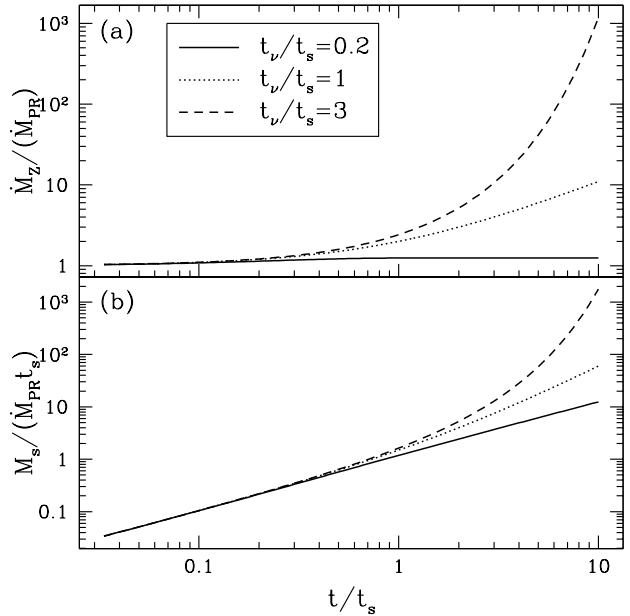}
\caption{
Time evolution of (a) $\dot M_Z$ and (b) total sublimated 
mass $M_s$. Evolutionary tracks are shown for different values
of the feedback parameter $t_\nu/t_s\equiv {\cal F}$ and 
demonstrate runaway behavior for $t_\nu/t_s >1$ and saturation 
of $\dot M_Z$ for $t_\nu/t_s <1$. Note that for $t\lesssim t_s$
the behavior of both $\dot M_Z$ and $\dot M_s$ is  
independent of  $t_\nu/t_s$.
\label{fig:f2}}
\end{figure}

The behavior of $\dot M_Z$ and $M_s$ is shown in Figure 
\ref{fig:f2} for different values of $t_s/t_\nu$, which clearly
demonstrates that for $t\lesssim t_s$ (i.e. as long as 
$M_s\lesssim\dot M_{PR}t_s$) debris disk evolution is 
insensitive to $t_s/t_\nu$ and $\dot M_Z\approx \dot M_{PR}$. 
This is because initially the 
disk of metallic gas is not dense enough for the drag it 
produces on the debris disk to compete with the PR drag ---
it takes certain time to accumulate enough gas mass by particle 
sublimation. 

This means, in particular, that if the debris disk starts with 
mass $M_d$, which is lower than the critical mass $\dot M_{PR}t_s$, 
then its evolution is determined only by the PR drag and coupling
to the gaseous disk is never effective --- the debris disk is
eroded before gas disks grows massive enough. Lifetime of the
debris disk is then 
\ba
t_{PR}=\frac{M_d}{\dot M_{PR}}\approx 3~\mbox{Myr}~
\frac{M_d}{10^{22}\mbox{g}}
\left(\frac{\dot M_{PR}}{10^{8}\mbox{g s}^{-1}}\right)^{-1},
\label{eq:t_PR}
\ea
independent of the relation between $t_s$ and $t_\nu$.

In the opposite case of a massive initial disk --- 
$M_d\gtrsim \dot M_{PR}t_s$ --- evolution does depend on
$t_s/t_\nu$. Whenever $t_\nu\lesssim t_s$ 
the action of viscosity is so effective at removing gas released 
by particle sublimation at $R_s$ that the gas does not 
accumulate there. Then $\Sigma_g$ simply saturates at 
the constant low level, accretion rate tends to $\dot M_Z\approx 
\dot M_{PR}(1-t_\nu/t_s)^{-1}\sim \dot M_{PR}$ for 
$t\gtrsim t_\nu$, and disk gets exhausted on timescale $t_{PR}$
given by equation (\ref{eq:t_PR}). This evolutionary path 
clearly cannot explain the highest observed values of $\dot M_Z$. 

However, in the case of a massive disk and $t_\nu\gtrsim t_s$ 
sublimation supplies gas to the $r\sim R_s$ region faster 
than viscous diffusion removes it, and $\Sigma_g$ grows 
exponentially on timescale $\sim t_s$
(as long as particle disk has enough mass to provide the
source). This means that $\dot M_Z$ also increases exponentially 
for $t\gtrsim t_s$ (see Figure \ref{fig:f2}). This runaway 
stops and gaseous disk 
wanes only when the mass of particulate disk $M_d$ is exhausted. 
The latter happens on the {\it runaway timescale} 
\ba
t_{\rm run}\approx t_s\left(1-\frac{t_s}{t_\nu}\right)^{-1}
\ln\left(\frac{M_d}{\dot M_{PR}t_s}\right),
\label{eq:t_end}
\ea
determined by the condition 
$M_s(t_{\rm run})\equiv\int^{t_{\rm run}}\dot M_Z dt\approx M_d$. 
Clearly, $t_{\rm run}\ll t_{PR}$ if $M_d/(\dot M_{PR}t_s)\gg 1$.

Thus, the ratio
\ba
{\cal F}\equiv\frac{t_\nu}{t_s}=\frac{4AR_s}{\alpha c_s^2}
\label{eq:feedback}
\ea
is the critical parameter determining the strength of positive 
feedback for massive disks: if ${\cal F}>1$ a particulate disk gets  
rather rapidly (within several $t_s$) converted into 
metallic gas at $R_s$, with its subsequent viscous accretion onto 
the WD on timescale of several $t_\nu$. 
Whereas for ${\cal F}<1$ feedback is not strong enough to
reinforce gas supply at $R_s$ and disk slowly evolves on timescale 
$t_{PR}$ due to PR drag. As equation (\ref{eq:feedback}) shows, 
runaway evolution and high $\dot M_Z$ 
require (1) strong coupling between the gaseous and particulate 
disks and (2) low viscosity.


\section{Aerodynamic coupling.}
\label{sect:aero}


Relative azimuthal motion between gaseous and particulate 
disks at speed $v_{\varphi}$ induces {\it aerodynamic drag}
between them. Drag produces force per unit area of the disk 
(Schlichting 1979)
\ba
f_\varphi=\mbox{Re}_\star^{-1}\rho_g v_{\varphi}^2,
\label{eq:aerodrag}
\ea
where $\rho_g=\Omega_K\Sigma_g/c_s$ is the gas density and 
Re$_\star^{-1}$ is a proportionality constant. 
There is a significant spread of opinions 
regarding the value of Re$_\star$ characterizing drag by the 
turbulent flow on a smooth solid plate, with numbers ranging 
between Re$_\star\approx 20$ (Dobrovolskis \etal 1999) to 
Re$_\star\approx 500$ (Goldreich \& Ward 1973). It is also
likely that a smooth plate approximation underestimates drag 
(overestimates the value of Re$_\star$) for the particulate 
debris disk, which does not have continuous surface 
and may interact with gas more like a rough plate (Schlichting 
1979) or even as a combination of individual particles, in 
which case smaller Re$_\star$ is more appropriate.

According to equation (\ref{eq:aerodrag}) aerodynamic 
drag force can be written in the form $f=A\Sigma_g$ with 
\ba
A=\frac{\Omega_K v_{\varphi}^2}{\mbox{Re}_\star c_s}=
\frac{\eta^2}{{\rm Re}_\star}\frac{c_s^3}{\Omega_K r^2},
\label{eq:A_a}
\ea
where equation (\ref{eq:v_rel}) was used. From equation 
(\ref{eq:tau_s}) sublimation time at $r=R_s$ is 
\ba
t_s=\frac{{\rm Re}_\star}{4\eta^2}\frac{G M_\star}{c_s^3}\approx
10^3\mbox{yr}\frac{{\rm Re}_\star}{\eta^2}\frac{M_{\star,1}}
{c_{s,1}^3},
\label{eq:t_s_a}
\ea
where $c_{s,1}\equiv c_s/(1$ km s$^{-1})$. The critical mass 
separating low- and high-mass disks (see \S\ref{sect:coupled}) 
is 
\ba
\dot M_{PR}t_s\approx 3\times 10^{18}\mbox{g}~
\frac{{\rm Re}_\star\phi_r}{\eta^2}\frac{M_{\star,1}}
{c_{s,1}^3}\frac{L_\star}
{10^{-3}L_\odot}\frac{20}{R_s/R_\star}.
\label{eq:crit_mass}
\ea
The feedback parameter for aerodynamic coupling is
\ba
{\cal F}=\frac{4\eta^2}{\alpha{\rm Re}_\star}
\frac{c_s}{\Omega_K R_s}\approx \frac{2\times 10^{-3}}{\alpha}
\frac{\eta^2}{\mbox{Re}_\star}
\left(\frac{T_{g,3}}{M_{\star,1}}
\frac{R_s}{0.2~\mbox{R}_\odot}\right)^{1/2}
\label{eq:F_a}
\ea
and depends on both the viscosity parameter $\alpha$ and the
strength of aerodynamic coupling, parameterized by Re$_\star$. 

Equation (\ref{eq:F_a}) shows that runaway evolution with 
${\cal F}\gtrsim 1$ requires rather low viscosity in the 
gaseous disk, at the level of $\alpha\sim 10^{-3}-10^{-4}$,
and for higher Re$_\star$ smaller $\alpha$ is needed. 
Viscosity is most likely provided by the magnetorotational 
instability (MRI), which requires certain level of
ionization to operate effectively. In the ideal MHD limit 
simulations with no net flux typically produce $\alpha\sim 
10^{-2}$ (Hawley \etal 1995). However, in our scenario 
gaseous disk exists in immediate contact with the
particulate disk, which is observationally known to contain 
a population of micron size dust grains (Jura \etal 2009). 
MRI-driven turbulence will mix some of this fine dust with the gas, 
lowering ionization fraction (small grains have large surface 
area and are very efficient charge absorbers), and giving 
rise to the non-ideal MHD effects, e.g via the increased 
resistivity (Balbus 2009). The latter are
known from simulations to decrease effective $\alpha$ 
substantially, down to $\alpha\sim 10^{-4}$ or lower 
(Fleming \etal 2000; Bai \& Stone, in preparation).
As a result, it is conceivable that $\alpha$ in our 
model can be much lower than in the ideal MHD limit 
of MRI.

Thus, presence of dusty debris disk in close contact with the 
gaseous disk can quite naturally lengthen viscous timescale 
$t_\nu$ and facilitate gas accumulation, making possible 
runaway evolution due to aerodynamic drag. We note that 
after the particulate disk completely sublimates   
the value of $\alpha$ in the resultant gaseous disk 
will go up since the source of small dust particles 
lowering ionization has disappeared. Closer to the WD, in 
the dust-free region $r<R_s$, ionization and $\alpha$ can 
{\it always} be higher than in the external disk. Faster 
viscous evolution in this region (resulting in lower 
$\Sigma_g$) may explain lack of line emission from the 
inner part of the gas disk (G\"ansicke \etal 
2006, 2007, 2008; Melis \etal 2010).


\section{Discussion.}
\label{sect:disc}


Our calculations clearly demonstrate that compact debris disks 
around WDs are self-destructive whenever there is strong coupling 
between the gaseous and solid components. Adopting for illustration 
Re$_\star=20$ and setting all other dimensionless constants
to unity we find from equations (\ref{eq:t_s_a})-(\ref{eq:F_a}) 
$t_s\approx 2\times 10^4$ yr, $\dot M_{PR}t_s\approx 
6\times 10^{19}$ g and that $\alpha\lesssim 10^{-4}$ is 
needed for ${\cal F}>1$. 

Then the typical lifetime 
of a massive disk with $M_d=10^{22}$ g (mass of a 200 km asteroid)
in the runaway scenario (if the viscosity is low) is on the order 
of several sublimation timescales, i.e. about $10^5$ yr. According
to equations (\ref{eq:MZsol}) and (\ref{eq:t_end}) in this case 
the {\it maximum} $\dot M_Z$ for this $M_d$, achieved right 
before the debris disk completely disappears, is 
max$(\dot M_Z)\approx M_d/t_s\sim 10^{10}$ g s$^{-1}$.
These numbers agree with observational inferences 
(Farihi \etal 2009, 2010) quite well. 

However, if positive feedback is not strong enough 
to drive runaway (e.g. because $\alpha> 10^{-4}$ or 
higher Re$_\star$), then $\dot M_Z\sim \dot M_{PR}$ 
and it takes 3 Myr for the same disk to be exhausted, 
as equation (\ref{eq:t_PR}) demonstrates. In this 
case gas surface density at $R_s$ saturates at the 
low level $\Sigma_g\sim \dot M_{PR}t_s/(\pi R_s^2)\sim 0.1$ 
g cm$^{-2}$, see equation (\ref{eq:sol}). 

Equation (\ref{eq:t_PR}) also shows that a 
low-mass disk with $M_d=10^{19}$ g (mass of a 20 km asteroid) 
is destroyed by PR drag very rapidly, within 
several thousand years (as long as the disk is optically thick
to incoming stellar radiation, see R11). 

Our model of debris disk evolution naturally explains coexisting
gaseous and particulate debris disks reported in Melis \etal 
(2010). Systems with reported IR excesses but lacking emission 
lines of high-Z elements in gas phase may simply possess lower 
mass gaseous disks, which have not had enough time to develop 
by sublimation of solids at $R_s$. Compositional variations between 
different WDs may also explain such systems.

Calculations presented in this work are rather simple and local 
in nature. We studied only one coupling mechanism --- 
aerodynamic drag, while other possibilities may be available 
as well (e.g. induction interaction (Drell \etal 1965; 
Gurevich \etal 1978) between the MRI-generated B-field in the 
gaseous disk and the debris disk particles). Also, here we 
did not consider low surface density disks
(we use only the thin plate approximation), non-trivial initial 
radial distribution of debris surface density, fate of angular 
momentum lost by debris disk and absorbed by gas disk, and so on.
Future global models of coupled evolution of gaseous 
and particulate debris disks (Bochkarev \& Rafikov, in preparation) 
will take these issues into account to provide a more accurate 
description of WD pollution with circumstellar high-Z material.

\acknowledgements

RRR is grateful to Xue-Ning Bai and Konstantin Bochkarev 
for useful discussions. The financial support of this work 
is provided by the Sloan Foundation, NASA via grant NNX08AH87G, and 
NSF via grant AST-0908269.




\end{document}